\documentstyle{article}

\newcommand{\bea}{\begin{eqnarray}}
\newcommand{\eea}{\end{eqnarray}}
\newcommand{\beq}{\begin{equation}}
\newcommand{\eeq}{\end{equation}}

\def\gsim{\mathrel{\raise.3ex\hbox{$>$\kern- .75em \lower1ex\hbox{$\sim$}}}}

\newfont{\cms}{cmss8 scaled 1440}

\def\/{\over}
\begin{document}

\parindent=1 em
\frenchspacing

\begin{center} {\LARGE \bf Shaping an ultracold atomic soliton in a travelling
 wave laser beam}
\\[1cm]
{\bf 
Markus Holzmann\footnote{Present adress: Laboratoire Kastler-Brossel and
Laboratoire de Physique Statistique; Ecole Normale Sup\'erieure;\\
\phantom{aaa} 24, rue Lhomond; F-75005 Paris; e-mail: holzmann@lps.ens.fr},
 and J\"urgen Audretsch\footnote{e-mail: Juergen.Audretsch@uni-konstanz.de}
\\[0.3cm]
\normalsize \it Fakult\"at f\"ur Physik der Universit\"at Konstanz\\
\normalsize \it Postfach 5560 M 674, D-78434 Konstanz, Germany}
\vspace{0.2cm}
\end{center}
\begin{abstract}
An ultracold wave packet of bosonic atoms loaded into a travelling laser wave may form a many-atom soliton.
This is disturbed by a homogeneous force field, for example by the inevitable gravitation.
The wave packet is accelerated and therefore the laser frequency appears to be chirped in the rest frame of the atoms.
We derive the effective nonlinear Schr\"odinger equation. It shows a time dependent nonlinearity coefficient which amounts to a damping or antidamping, respectively.
The accelerated packet solution remains a soliton which changes its shape adiabatically. Similarly, an active shaping can be obtained in the force-free case by chirping the laser frequency 
thus representing a way of coherent control of the soliton form.
The experimental consequences are discussed.
to be published in Europhys. Lett.

\end{abstract}

\vspace{0.3cm}
After the recent realizations of Bose-Einstein condensation in ultracold and dilute gases [1-4], it may be expected for the near future that a new source of atoms will be available which provides atoms of high flux and coherence. 
The output coupler of such an atom laser is based on atom-light interaction \cite{Burnett2}: Coherent coupling of atoms in two distinct hyperfine states by microwaves is used in \cite{Boser} to transfer the Bose condensate from a trapped to an untrapped state.  A significant consequence of the quantum field theory of ultracold bosons interacting with photons \cite{Lewenstein,Zhang}
is that two-body interactions like the dipole-dipole interaction lead to nonlinear effects in the propagation of wave packets.
This opens a new exciting field of research in atom optics called nonlinear atom optics. It has been worked out by several authors \cite{Walls/soliton,Meystre1,Meystre2}
that due to this nonlinearity an ultracold atomic wave packet 
may form a many-atom soliton guided by a travelling or standing laser mode of constant 
frequency. Thus the atoms are immune to wave packet spreading if no additional forces are
present.
Evidently the realization of such solitons is of great
importance for atom interferometry and other domains of quantum optics.

The considerations mentioned above are in some sense incomplete.
Atoms are massive and therefore necessarily subject to the influence of the gravitational field of the earth.
This is one of the main differences between nonlinear photon optics and nonlinear atom optics. In the set up of \cite{Boser} it is the acceleration due
to gravitation which gives the outcoupled atoms a distinct direction.
Will this destroy the soliton character of the matter wave packet, so that in fact 
the set up sketched above is of limited use?
In any case a reshaping of the wavepacket is to be expected.
Let us therefore look at the same problem from a different point of view:
Consider an unaccelerated ultracold many-boson soliton (no force present) and
study the influence of a frequency chirping of the laser wave. Will this allow us to manipulate the packet envelope so that a coherent control of the soliton
shape becomes possible? 
It is the aim of this paper to show that this can indeed be done.
To do so we have in a first step to reexamine all the underlying approximations
used to derive soliton solutions in the force-free (or constant laser frequency)
case.
We will thereby assume the reader to be familiar with \cite{Walls/soliton}.

We discuss an ideal gas of identical two-level atoms interacting with the 
vacuum electromagnetic field and a travelling laser wave.
The gas is assumed to be composed of ultracold Bose atoms, which can be treated as a vector quantum field with two components $\hat{\psi}_g$ and $\hat{\psi}_e$
corresponding to the internal ground state and the excited state of the atoms,
respectively. The internal energy difference is $\hbar \omega_a$.
Due to the coupling to the electromagnetic vacuum (virtual) photon exchanges between different ultracold atoms are possible.
After the elemination of the vacuum field these photon exchanges result
in a long range interatomic correlation which is responsible for the existence of atomic solitons.
Following \cite{Walls/soliton} we propose a scheme where ultracold atoms are loaded into a travelling laser beam with frequency $\omega_l$, which propagates in the z direction.
But in contrast to \cite{Walls/soliton} the additional influence of a homogeneous force field parallel to the z direction is explicitly taken into account. This leads in the inertial reference frame to additional potential terms $-m a z \hat{\psi}_{g/e}$ in the coupled differential equations for the ground and excited field components of the atomic gas whereby $a$ may be positive or negative.

Let us first note the equivalence of the influence of a homogeneous
force field to a frequency chirping.
The narrow atomic wave packet is assumed to propagate collinearly
within the laser beam. The following transformations
\bea
\hat{\psi}_g({\bf r},t) & = & \hat{\phi}_g({\bf r},t)
 e^{ik_G z} e^{-i \alpha (z, t)},
\label{AA1} 
\\
\hat{\psi}_e({\bf r},t) & = & \hat{\phi}_e({\bf r},t)
e^{i(k_G+k_l) z- i\omega_l t} e^{-i \alpha (z, t)},
\label{AA2}
\\
{\widetilde \Omega}^{(+)} & = &\Omega^{(+)}(x,y) e^{ik_lz- i \omega_l t }
\label{AA3}
\eea
where $k_G$ is the initial wave vector of the center of the atomic beam and 
where
\beq
 \alpha (z, t) = {1 \over \hbar} \left( {\hbar^2 k_G^2 \over{2 m}} t - mazt + {ak_G  t^2 \over 2} + {ma^2 t^3 \over 6} \right)
\eeq
corresponds to the total kinetic energy of the atomic beam, yield the equations of motion for the atomic field envelopes $\hat{\phi}_{g/e}({\bf r},t)$ and the Rabi frequency of the laser field $\Omega^{(+)}({\bf r})$.
These equations are of such a nature that after having changed to the co-moving frame of the accelerated atoms by
\beq
z \rightarrow z - \hbar k_G t/m - \frac{1}{2} a t^2,
\label{AA5}
\eeq
the effect of the homogeneous force field is reduced to a chirping of the laser frequency.
This means that after the transformations (\ref{AA1})-(\ref{AA5}) we obtain the
same equations of motion for the envelope fields as in the forcefree case
(\cite{Walls/soliton}, group velocity $v_g=\hbar k_g/m= 0$)
but with a time dependent effective detuning $\Delta(t)$ due to the Doppler shifted laser detuning
\beq
\Delta(t) = \omega_l - \omega_a -  \omega_r - k_l (\hbar k_G/m +at)
\label{AA6}
\eeq
where $\omega_r = \hbar k_l^2 /2m$ is the recoil frequency. We introduce $\Delta_0 = \Delta(a=0)$ for later use.

To avoid the influence of spontaneous emission,
which induces inelastic processes, we assume a large detuning. The 
probability of light scattering by single atoms, proportional to $(\Omega/
\Delta)^2$, needs to be sufficiently small. However, for negative detuning
the mean field interaction between atoms can be significantly changed
without excessive recoil or photoassociation losses due
to the presence of vibrational quasimolecular states \cite{Fedichev}.

In that region far from atomic resonance (that holds as long as $|\Omega / \Delta(t)|\ll 1$) we can adiabatically eliminate the excited state component
 up to the order $\Delta^{-2}(t)$ and obtain the nonlinear Schr\"odinger equation of Ref.
\cite{Walls/soliton} but with the time dependent detuning $\Delta(t)$.
This adiabatic elemination is possible as long as the
relative change of the detuning due to its time dependence is small,
that applies for times $t\ll T$ with $T=|\Delta_0/k_l a|$. The resulting equation 
of motion for the ground state field operator of the atoms contains
an effective single particle potential $V(x,y) \sim
\Omega(x,y)/\Delta(t)$ (the usual dipole potential from the laser field)
and a nonlinear part $\sim \Delta^{-2}(t)$ due to the two-body dipole-dipole 
interaction.

Essentially all $N$ ultracold Bose atoms occupy the same state.
This justifies the time-dependent Hartree approximation for the coherent evolution of the ultracold
atoms \cite{Meystre1,Wright}:
The $N$-atom wavefunction $\Phi({\bf r}_1,{\bf r}_2, \dots , {\bf r}_N;t)$
is written as a product of $N$ single atom wavefunctions $\Phi({\bf r}_i,t)$
\bea
\Phi({\bf r}_1,{\bf r}_2, \dots , {\bf r}_N;t) \simeq \prod_{i=1}^N \Phi({\bf r}_i,t).
\label{HN1}
\eea
 
To be consistent  with the
Hartree approximation (\ref{HN1}) we neglect dissipative processes
due to spontaneous emissions and laser driven collisions between
the ultracold atoms. Spontaneous emission can be ignored
for a highly detuned laser ($|\Delta(t)|\gg \gamma$, where $\gamma$ 
is the spontaneous emission rate) as shown numerically in  \cite{Spont}
for a comparable situation, whereas the effect of binary collisions
depends strongly on the laser detuning \cite{Burnett}.
For a red detuned laser we must assure that the laser does not excite a bound state level $\nu$, which can be built up due to the dipole-dipole potential between a ground and an excited atom. 
If the detuning of the laser is choosen between two vibrational levels with energy difference $\hbar \delta_{\nu}$, this condition is fulfilled
for $t<T_{\nu}$ with $T_{\nu} =|\delta_{\nu}/k_l a|$ if 
 the linewidths of the excited bound states $\gamma_{\nu}$ are small compared with $\delta_{\nu}$ ($\gamma_{\nu}/\delta_{\nu} \ll 1$).
Otherwise the laser driven collision rate is increased, the colliding pairs get a higher momentum than the other atoms and cannot be described by the Hartree ansatz. 

Furthermore we replace the true dipole-dipole 
interaction by a contact potential \cite{Meystre1}.
This is not allowed in general and has been criticised \cite{Castin},
but should be reasonable for ultracold atoms. Since the range of the
dipole-dipole interaction is of the order of an optical wavelength
$\lambda = 2 \pi /k_l$, the description of the full potential by its first moment is expected to be valid for the long thermal de Broglie wavelenth of the ultracold atoms $\lambda_{dB} \gg \lambda$. 
In practice the spatial orientation of the atomic dipole moment ${\bf d}$ is uncertain, so that we must take the mean value over the orientation of the dipoles in the dipole-dipole interaction. With the averaged dipole-dipole interaction we get a positive correlation volume $V_c = 2 \pi /8 k_l^3$.

Thereby we obtain as an intermediate result a nonlinear Schr\"odinger equation for the effective single particle states of the ground state atoms:
\bea
i\hbar {\partial \Phi \over {\partial t}}
 =  - {\hbar^2 {\bf \nabla}^2 \over {2m}} \Phi
  + {\hbar |\Omega(x,y)|^2 \over {4 \Delta(t)}} \Phi  - {N \hbar \gamma V_c |\Omega(x,y)|^2 \over{2 \Delta^2(t)}}
  |\Phi|^2 \Phi. 
\label{HN2} 
\eea
The possibility of manipulationg the value and the sign of the scattering 
length, corresponding to the mean field interatomic interaction, is also
studied in \cite{Fedichev}, where the neglect of the dissipative terms and
the contact potential approximation are justified in another approach.

Let us now try to find a solution of the nonlinear Schr\"odinger equation
with a propagation of the atoms guided by the laser beam.
Since the single particle potential in (\ref{HN2}) is independent of the longitudinal
coordinate z, we split the wave function in a transverse  and a longitudinal part
by a separation ansatz
$\Phi({\bf r},t)=  u(x,y,t)\varphi(z,t)$. 
The transverse motion of the single particle wave function is then determined by
\bea
i\hbar {\partial u(x,y,t) \over {\partial t}}
 =  - {\hbar^2 {\bf \nabla_{\bot}}^2 \over {2m}} u(x,y,t)
  - {\hbar |\Omega(x,y)|^2 \over {4 |\Delta(t)}|} u(x,y,t).
\label{TWF1}
\eea
It contains a potential term which depends on the transverse laser profile
$\Omega(x,y)$ and decreases with increasing detuning.
For an appropriate profile, which may for example be a Gaussian, one can
realize for the transverse wave function bounded energy eigenmodes
$u_0(x,y,t)$.
We assume the atomic ensemble to be in one of these eigenmodes. 
Due to the adiabatic theorem it will stay in this eigenmode
collecting only an additional dynamical phase factor which will not
be relevant as long as $\epsilon T \gg \hbar$ is fullfilled
whereby $\epsilon$ is the energy difference to the next energy eigenstate
and $T$ is the typical time scale for the time dependence 
of the single particle potential introduced above.
Turning to the particular case of an external force acting, it is worth
to note that if the force $m{\bf a}$ is not collinear to the laser momentum
then respective components in x or y direction result only in a displacement
of the atomic center of mass in these directions, which can be compensated by an appropriate shape of the laser profile. 

In addition to this transversal confinement it is the longitudinal behaviour
which is of central importance for the shaping of ultracold atomic solitons
and their stability in external force fields.
We then finally obtain the following nonlinear Schr\"odinger equation in the co-moving accelerated frame
\beq
i\hbar \frac{\partial \varphi}{\partial t} +
\frac{\hbar^2}{2m}  \frac{\partial^2 \varphi}{\partial
z^2}
+ \hbar \kappa(t) N|\varphi|^2 \varphi = 0
\label{S3} 
\eeq
where
\beq
\kappa(t) = \frac{\gamma V_c}{2\Delta(t)^2}
\int dx \! \int dy \, |u_0(x,y,t)|^4 |\Omega(x,y)|^2 
\label{S2} 
\eeq
is the nonlinearity coefficient which has become time dependent because of the influence of the force field. It corresponds to the Kerr-type nonlinear susceptibility in nonlinear optics. The coefficient demonstrates how with increasing effective detuning $|\Delta(t)|$ the dipole-dipole interaction is switched off and the nonlinearity in (\ref{S3}) disappears.

The important consequence of acceleration or chirping is that 
the nonlinearity coefficient $\kappa$ is now time dependent. An adiabatic treatment is a priori not justified in a nonlinear equation.
Therefore it is very useful to make the following transformation to separate the time dependence from the nonlinearity
\beq
\varphi = \left( \frac{2m \kappa(t) N }{\hbar} \right)^{-1/2} \widetilde{\varphi}.
\label{DS1} 
\eeq
We obtain
\beq
i \frac{2m}{\hbar} \frac{\partial \widetilde{\varphi}}{\partial t} +
  \frac{\partial^2 \widetilde{\varphi}}{\partial
z^2}
+ |\widetilde{\varphi}|^2 \widetilde{\varphi} + i\frac{m}{\hbar}\Gamma(t) \widetilde{\varphi} = 0.
\label{DS2}
\eeq
This reveals the underlying physics: The time dependence of $\kappa$ gives rise to an additional damping coefficient 
\beq
\Gamma(t) = -  \frac{\dot{\kappa(t)}}{\kappa(t)}
\label{DS3} 
\eeq
which is of the order $T^{-1}$. 
An increasing nonlinearity coefficient $\kappa(t)$ reflects according to (\ref{S3}) a growing longitudinal confinement due to the laser induced nonlinearity, and leads to antidamping. This is the case for the red detuned laser field propagating in reverse direction of the force field ($a<0$).
If both directions agree ($a>0$) damping occurs.

For times $t$ with $|\Gamma| t \ll 1$ that is for $t\ll T$ the damping term in (\ref{DS2}) may be treated perturbationally \cite{Lamb}.
Returning to the longitudinal part of the wave packet envelope we get (for
$a\neq 0$ still with reference to the co-moving accelerated frame) a normalized soliton solutions of the following form:
\beq
\varphi_s(z,t) = \sqrt{ \frac{m \kappa(t) N}{4\hbar} } {\rm sech}[\frac{m \kappa(t) N}{2 \hbar}z] \exp\left[-i\frac{m\kappa_0^2 N^2}{4 \hbar} \int_0^t dt' \, \left(\frac{\kappa_0}{\kappa(t')}\right)^4\right] ,
\label{DS4} 
\eeq
with $\kappa_0 = \kappa(t=0)$. This is the intended result.
Under the influence of a homogeneous force field acting on the atoms
or for a linearly chirped laser, the soliton changes its shape adiabatically
in time due to the time dependence of the nonlinearity coefficient $\kappa(t)$.
This opens the possibility of a coherent control of the shape of the 
 soliton. The relative change of the width $W(t)=\sqrt{2}\hbar/{m \kappa(t) N}$
is constant $\dot{W}/W = - 2 T^{-1} {\rm sign} (a \Delta_0)$.
The related relative change of the height of the soliton is given by the inverse expression.
The damping and antidamping mentioned above can be read off directly.
If in the case of acceleration we go back to the inertial reference frame
we see that the pulse maximum is also accelerated due to the homogeneous force.

Let us now turn to the experimental realization and estimate the influence of the earth gravitational field.
 We consider the $[3S_{1/2} \rightarrow 3P_{1/2}]$ transition of sodium atoms with wavelength $589nm$ ($k_l = 10^{7}m^{-1}$) and a natural linewidth of $\gamma/2\pi = 10 MHz$.
 We assume a red detuning $\Delta_0$ of $100 GHz$ and for the Rabi frequency $\Omega_0 \sim \Delta_0/10$ so that the adiabatic elemination of the excited atom state is justified.
 To keep the relative change of the detuning small  and to justify our approximations, we have to restrict to times $t$ smaller than $T=|\Delta/k_l a| \sim 10^3 s$.
 In fact, with $\gamma_{\nu} = 20 MHz$ as the linewidth of molecular bound states
 and $\delta_{\nu} \simeq 10 GHz$ \cite{Burnett}, the cold collisions mentioned above can only
 be avoided during a smaller time up to approximately $T_{\nu}= |\delta_{\nu}/k_l a| \sim 100 s$
 since $\gamma_{\nu}/\delta_{\nu} \ll 1$, which is still a rather long travelling time for the atomic sample.
 To discuss the transversal confinement, we assume a Gaussian laser
 profile $\Omega(x,y) =\Omega_0 \exp[-(x^2+y^2)/2b_l^2]$ with a width $b_l$ of the laser beam
 much larger than the width of the atomic beam. Accordingly we obtain approximately harmonic
 oscillator eigenfunctions for the transverse atomic mode $u_0(x,y,t)$ with a width $b_a$
 of the fundamental eigenmode.
It can then be shown that the condition $\epsilon T \gg \hbar$ necessary for the adiabatic
 treatment of the transversal modes amounts to the condition $|k_l a /\Omega_0^2 |\ll (b_a/b_l)^2 \ll 1$
 which can always be fulfilled. For the experimental specification given above all approximations
 are therefore justified and our result (\ref{DS4}) may be applied.
To demonstrate how effective the confining nonlinearity still is,
 let us assume that the influence of the gravitational force field maintains for $T_{\nu}= 100 s$.
 The width and the height of the soliton is then changed only by $20\%$. This is compared with the
 spreading of an ordinary Schr\"odinger wave packet with inital width $\overline{W}_0$ under
 the same conditions: $\overline{W}(t) = (1 + (\omega_r t)^2 (k_l \overline{W}_0)^{-4})^{1/2}$.
 For $\overline{W}_0 k_l \sim 100$, which is consistent with the initial width for the soliton
 in our case ($N \sim 10^{5}$), the width of the Gaussian increases by about $40\%$ in $4*10^{-7}s$.
 In the time $T_{\nu}=100s$ the width has increased by a factor of $10^{8}$.

We turn to the positive aspect of an active manipulation of the soliton shape.
 For this the effective detuning may be changed by chirping of the laser frequency.
 If the resulting time dependence is linear, it can be described by an effective acceleration $\widetilde{a}$.
 Then, to obtain a change of the soliton width of $20\%$ in already $1s$ of chirping,
 an effective acceleration of $\widetilde{a}= 100 g$ is necessary.

We thank F. Burgbacher and K.-P. Marzlin for fruitful discussions. M.H. also acknowledges
 helpful discussions with Y. Castin and financial support by the Optik Zentrum Konstanz.

\end{document}